\documentclass[twocolumn]{aastex62}
\pdfoutput=1

\usepackage{amsmath}

\received{February 6, 2019}
\revised{April 29, 2019}
\accepted{May 1, 2019}
\submitjournal{ApJL}

\begin{document}

\title{Dust Polarization in Four Protoplanetary Disks at 3 mm: Further Evidence of Multiple Origins}

\author{Rachel E. Harrison}
\affiliation{Department of Astronomy, University of Illinois at Urbana-Champaign, 1002 W. Green St., Urbana, IL 61801, USA}
\author{Leslie W. Looney}
\affiliation{Department of Astronomy, University of Illinois at Urbana-Champaign, 1002 W. Green St., Urbana, IL 61801, USA}
\author{Ian W. Stephens}
\affiliation{Harvard-Smithsonian Center for Astrophysics, 60 Garden St., Cambridge, MA 02138, USA}
\author{Zhi-Yun Li}
\affiliation{Department of Astronomy, University of Virginia, Charlottesville, VA 22903, USA}
\author{Haifeng Yang}
\affiliation{Department of Astronomy, University of Virginia, Charlottesville, VA 22903, USA}
\affiliation{Institute for Advanced Study, Tsinghua University, Beijing, 100084, People’s Republic of China}
\author{Akimasa Kataoka}
\affiliation{National Astronomical Observatory of Japan, 2-21-1 Osawa, Mitaka, Tokyo 181-8588, Japan}
\author{Robert J. Harris}
\affiliation{Department of Astronomy, University of Illinois at Urbana-Champaign, 1002 W. Green St., Urbana, IL 61801, USA}
\affiliation{National Center for Supercomputing Applications, University of Illinois at Urbana-Champaign, 1205 W Clark St, Urbana, IL 61801, USA}
\author{Woojin Kwon}
\affiliation{Korea Astronomy and Space Science Institute (KASI), 776 Daedeokdae-ro, Yuseong-gu, Daejeon 34055, Republic of Korea}
\affiliation{University of Science and Technology, Korea (UST), 217 Gajeong-ro, Yuseong-gu, Daejeon 34113, Republic of Korea}
\author{Takayuki Muto}
\affiliation{Division of Liberal Arts, Kogakuin University, 1-24-2 Nishi-Shinjuku, Shinijuku-ku, Tokyo, 163-8677, Japan}
\author{Munetake Momose}
\affiliation{College of Science, Ibaraki University, 2-1-1 Bunkyo, Mito, Ibaraki 310-8512, Japan}

\begin{abstract}

We present polarimetric observations of four Class II protoplanetary disks (DG Tau, Haro 6-13, RY Tau, and MWC 480) taken with the Atacama Large Millimeter/submillimeter Array (ALMA) at 3 mm. The polarization morphologies observed fall into two distinct categories: azimuthal polarization (DG Tau and Haro 6-13) and polarization parallel to the disk minor axis (RY Tau and MWC 480). The mechanisms responsible for disk polarization at millimeter and submillimeter wavelengths are currently under debate. In this Letter, we investigate two mechanisms capable of producing polarized dust emission in disks: self-scattering and grain alignment to the radiation anisotropy. The polarization morphologies of DG Tau and Haro 6-13 are broadly consistent with that expected from radiation alignment (though radiative alignment still does not account for all of the features seen in these disks), while RY Tau and MWC 480 are more consistent with self-scattering. Such a variation in the polarized morphology may provide evidence of dust grain size differences between the sources.

\end{abstract}



\section{Introduction} \label{sec:intro}

One of the initial motivations for making resolved polarimetric observations of thermal emission from dust grains in circumstellar disks \citep[e.g.,][]{2014Stephens, rao14, segura-cox15} was to use magnetically aligned dust grains to infer field morphology and strength.
Magnetic fields are thought to play a crucial role in the accretion process in protostellar disks through magnetorotational instability \citep{1998RMP...70...1} or magnetic disk winds \citep{1982MNRAS...199...883}. Spinning dust grains in a magnetic field will align with their short axes parallel to the magnetic field due to radiative torques, producing polarization perpendicular to the magnetic field lines \citep[e.g.,][]{lazarian}. Possible evidence of grain alignment to the disk magnetic field has been found in the circumbinary disks BHB07-11 \citep{alves} and VLA 1623 \citep{sadavoy, harris18}, and the disk of HD 142527 \citep{ohashi}. However, in the past few years, two other mechanisms for producing dust continuum polarization have been discussed as more consistent with the observations: Rayleigh scattering \citep[e.g.,][]{kataoka15, yang16, yang17} and grain alignment to the radiation anisotropy \citep[e.g.,][]{kataoka, 2017ApJ...839...56}. 

Scattering is most efficient for grains about $\sim \lambda/2\pi$ in radius, where $\lambda$ is the observing wavelength \citep[e.g.,][]{kataoka15}. The polarization fraction produced by scattering is highly dependent on the observing wavelength for a given dust grain population. Therefore, determining the wavelength at which the scattering polarization peaks can provide a constraint on the dust grain size independent of the dust opacity index $\beta$ \citep{kataoka15, kataoka}. This independent estimate is useful because $\beta$ values can be affected by dust grain porosity \citep{kataoka14}, and because the derived value of beta (as inferred from an optically thin model) can be reduced by optically thick rings \citep{ricci}. Disk geometry affects the orientation and degree of polarization from scattering. In an optically thin disk, a higher inclination angle can induce a higher polarization fraction, and the polarization vectors will be more aligned with the minor axis of the disk \citep{yang16}. Polarized emission can also be produced when elongated dust grains align with their short axes along the direction of radiation anisotropy of the disk (generally outward). Larger grains ($>$~mm-sized) are not expected to align with the disk magnetic field, but are expected to be aligned by radiative torques perpendicular to the radiation anisotropy \citep{2017ApJ...839...56}. Polarization from radiatively aligned grains has a weaker wavelength dependence than that from self-scattering. Because the degree of polarization from radiative alignment depends on the projected shape of the grain, the observed polarization depends on the disk inclination and varies azimuthally \citep{yang18}.
Overall, multi-wavelength polarimetric observations of circumstellar disks provide a unique tool to probe dust and disk properties, including grain size and their distribution, in the disk.

HL Tau is the prototypical source for investigating how disk polarization changes with observing wavelength. Observations of HL Tau with the Atacama Large Millimeter/submillimeter
Array (ALMA) at 870 $\mu$m, 1.3 mm, and 3.1 mm revealed that the disk's polarization morphology changes rapidly with wavelength \citep{2017ApJ...851...55S}. At 870 $\mu$m, the polarization vectors were aligned parallel to the disk's minor axis. This is consistent with the pattern expected from Rayleigh scattering. The polarization vectors are azimuthally oriented at 3.1 mm. The azimuthally oriented polarization was initially thought to be evidence of radiative grain alignment, but as we will discuss, there are significant discrepancies between the polarization predicted by models and that observed at 3.1 mm \citep[see also][]{yang18}. Observations at 1.3 mm show a mix of the two morphologies. The dependence of polarization morphology on wavelength in HL Tau motivated the need for polarimetric observations of other circumstellar disks. Our study maps the polarization of four Class II disks at 3 mm: DG Tau, Haro 6-13, RY Tau, and MWC 480.

These sources are located in the Taurus Molecular Cloud; we have assumed a distance of 140 pc for all sources \citep[e.g.,][]{bacciotti18}. DG Tau has a protostar mass of 0.7 $M_\odot$ and an inclination angle of 32$^{\circ}$ \citep{2011AA...529A.105G}. A recent paper by \citet{bacciotti18} presented ALMA polarization observations of DG Tau at 870 $\mu$m. These observations showed an asymmetry in the polarized intensity along the disk minor axis. The polarization angles were oriented parallel to the minor axis near the disk center, but showed a more azimuthal orientation near the disk's edge. The  asymmetry seen in polarized intensity along the minor axis of the disk may indicate that the grains responsible for scattering at 870 $\mu$m have not settled to the disk midplane. Haro 6-13 has a protostar mass of 0.55 $M_\odot$ \citep{2011AA...529A.105G, kwon} and an inclination angle of 40$^\circ$ \citep{schaefer09}. MWC 480 has a protostar mass of 1.91 - 2.2 $M_\odot$ and an inclination angle of 36$^\circ$ \citep{2018arXiv181006044L}; the disk has a gap with a ring at a radius of 97.58 $\pm$ 0.08 au \citep[]{2018arXiv181006044L, hamidouche}. RY Tau has a protostar mass of 2.04 $M_\odot$ and an inclination angle of 65$^\circ$, with a ring at 18.19 $\pm$ 0.00 au \citep{2018arXiv181006044L}. The estimated $\beta$ values for these disks are 0.57 for DG Tau \citep{2011AA...529A.105G}, consistent with zero or a small positive value for Haro 6-13 \citep{kwon}, 0.86 for MWC 480 \citep{2011AA...529A.105G}, and 0.3 for RY Tau \citep{ricci}. 
%
%

\section{Observations}
The observations were taken with ALMA between 2017 November 29 and 2017 December 3, in ALMA configuration C43-7. The target sources were DL Tau, Haro 6-13, MWC 480, DG Tau, V982 Tau, and RY Tau. The observations were taken at a frequency range of 91.48–103.54 GHz (ALMA Band 3). J0522--3627 was the polarization calibrator, and J0510+1800 was the bandpass calibrator and flux calibrator. J0426+2327 was the phase calibrator for DL Tau, Haro 6-13, and DG Tau. J0512+2927 was the phase calibrator for MWC 480, J0403+2600 was the phase calibrator for V892 Tau, and J0438+3004 was the phase calibrator for RY Tau. Polarization was detected at or above the 3$\sigma$ level in DG Tau, Haro 6-13, MWC 480, and RY Tau, but not in DL Tau or V892 Tau. The 3-$\sigma$ upper limits on the polarization fraction in DL Tau and V892 Tau are 0.8\% and 0.4\%, respectively. DL Tau and V892 are not outliers from the other disks in terms of inclination angle or total intensity (see Table 1). The lack of polarized emission may have to do with the dust properties of these disks; e.g., they may not contain enough dust grains of the size necessary for the observed scattering at 3 mm. Multi-wavelength observations will be needed to determine why these sources lack polarized emission at 3 mm. 

The dataset was calibrated by data analysts at the North American ALMA Science Center. After this initial calibration, we performed three rounds of phase-only self-calibration on all Stokes parameters ($I$, $Q$, $U$, and $V$). We used the CASA task tclean with Briggs weighting and a robust parameter of 0.5. Cleaning thresholds were set based on the Stokes $I$ rms in non-self-calibrated images. The threshold for the first three iterations of tclean was set to 10 times the $I$ rms, and the threshold for the fourth iteration was set to three times the $I$ rms. The first iteration of gaincal used a solution interval equal to the scan length, the second iteration used a solution interval of 30.25 s, and the third iteration used a solution interval of 15 s. Polarization angle and intensity maps were produced from the Stokes $Q$ and $U$ data. The polarized intensity maps were debiased using the average noise value determined from the $Q$ and $U$ maps, an estimator used by e.g. \citet{wardle} and \citet{vidal16}: 

\begin{equation}
  P=\begin{cases}
    \sqrt{Q^2 + U^2 - \sigma^2} & \text{if $\sqrt{Q^2 + U^2} \geq \sigma$} \\
    0 & \text{otherwise}
  \end{cases}
\end{equation}

All images had an angular resolution between 0.2 and 0.3 arcsec. The uncertainty on absolute flux calibrations with ALMA is estimated at $\sim$10\%. ALMA's instrumental limit for a 3$\sigma$ detection of polarized emission is 0.1\% polarization for compact sources within one-third of the primary beam. The typical sensitivities of the Stokes $Q$ and $U$ images are ~20-40 $\mu$Jy/beam. For the rest of this Letter only statistical uncertainties are considered. 

\section{Results} \label{sec:results}

\begin{figure*}[ht]
\gridline{\fig{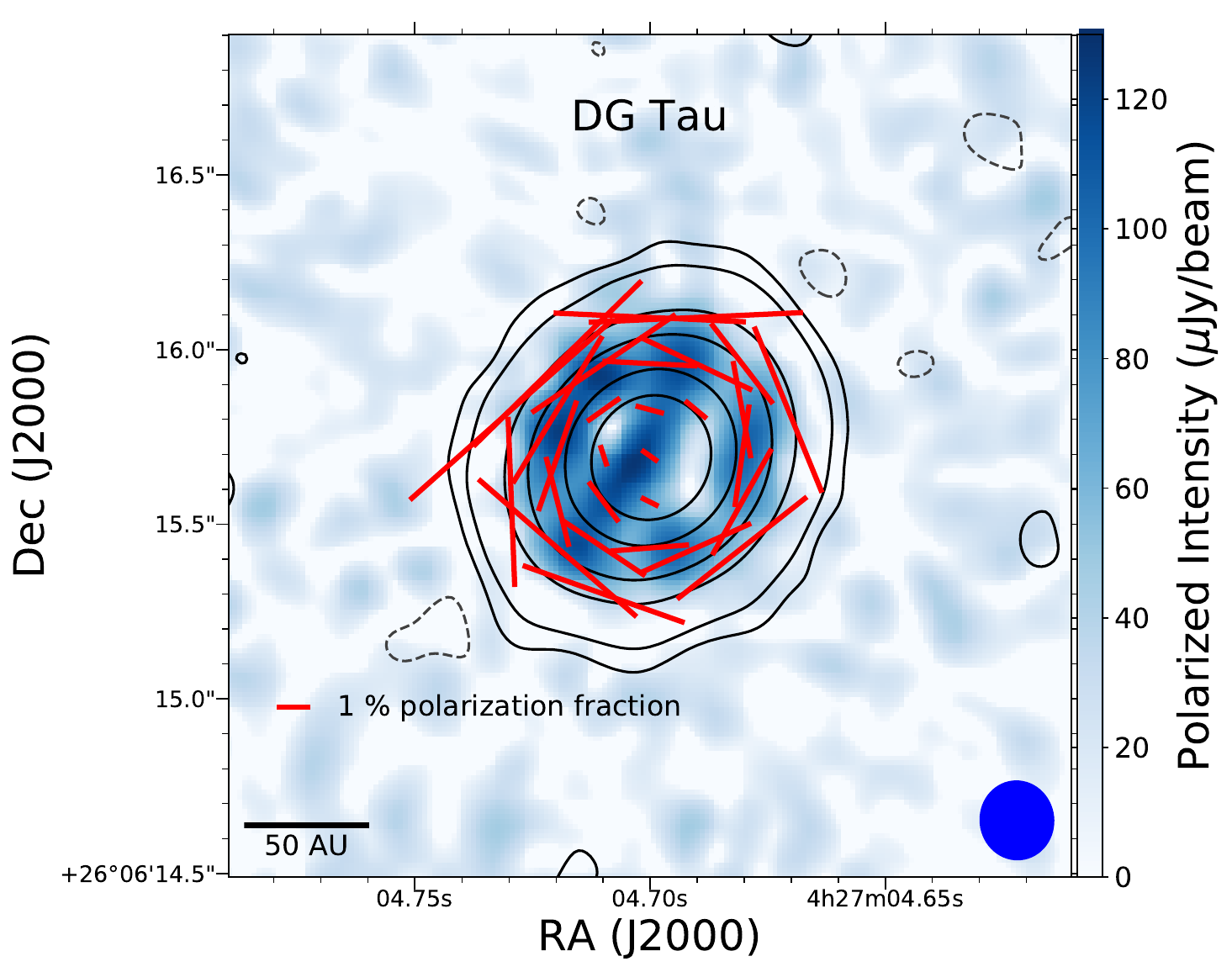}{0.45\textwidth}{(a)}
          \fig{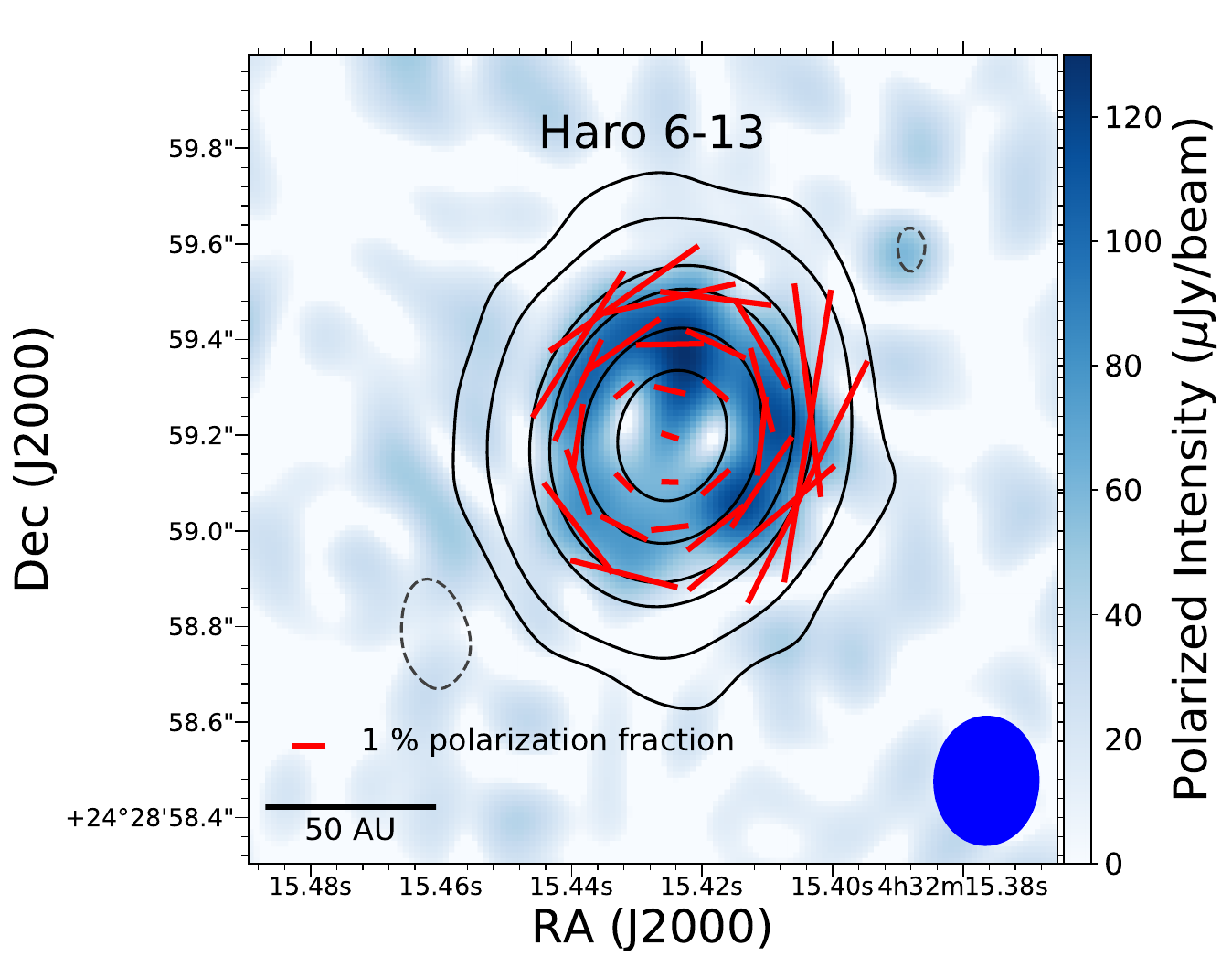}{0.45\textwidth}{(b)}}
\gridline{\fig{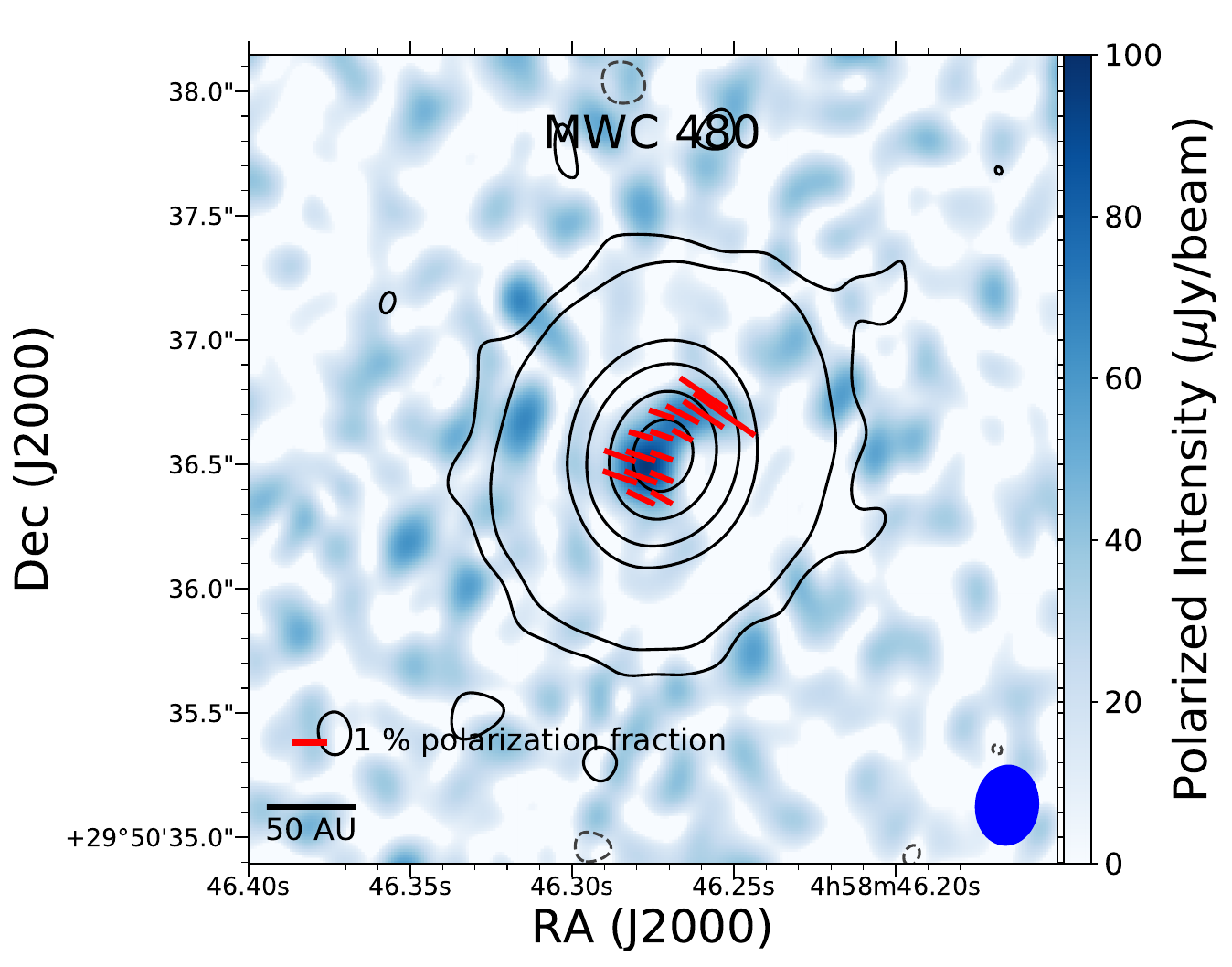}{0.45\textwidth}{(c)}
          \fig{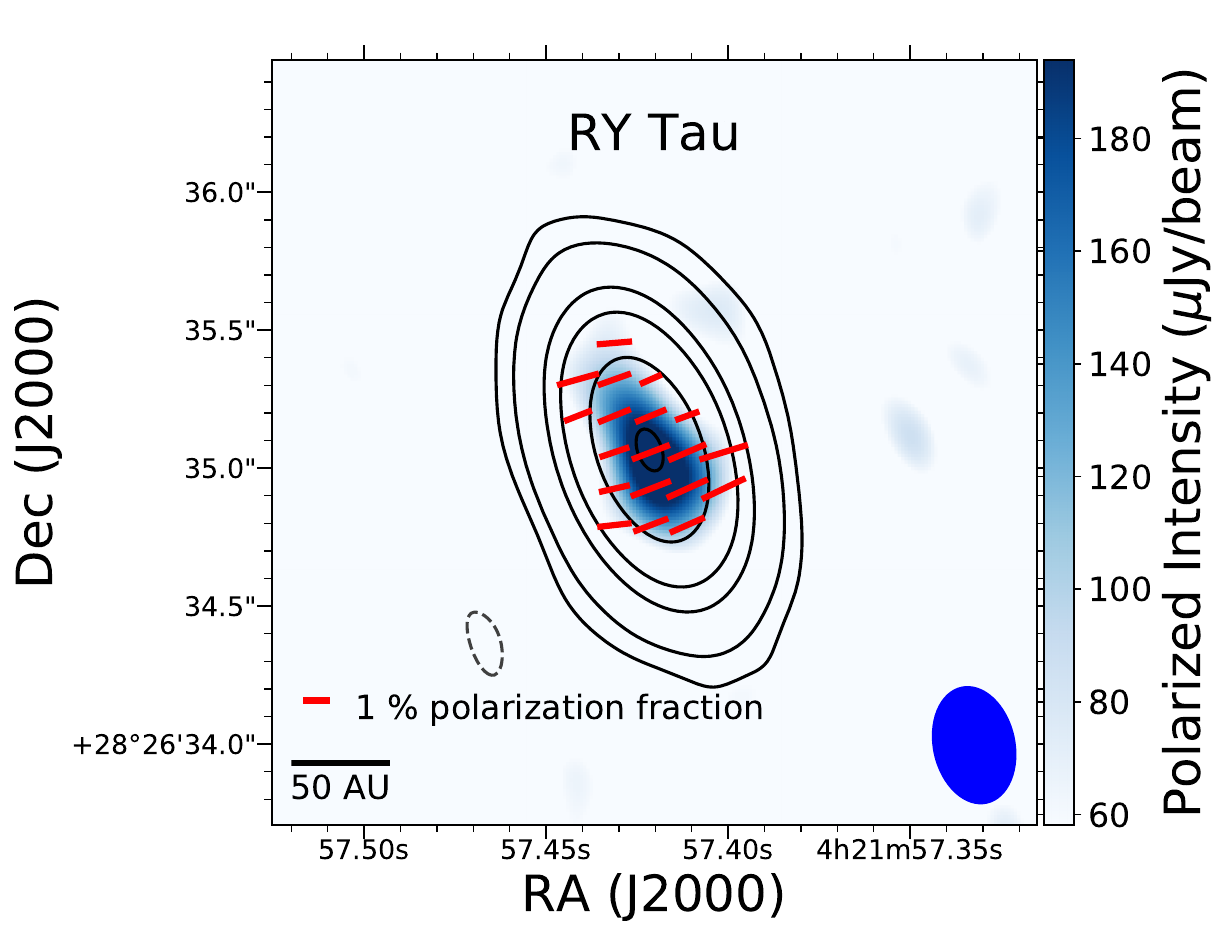}{0.45\textwidth}{(d)}}

\caption{Images of DG Tau, Haro 6-13, MWC 480, and RY Tau at 3 mm. The contours represent total intensity (Stokes $I$) of -3 (dashed), 3, 10, 50, 100, 250, and 500$\sigma$ levels, where $\sigma$ is listed in Table \ref{obs}.  The colormap represents polarized intensity with the scale on the right of each source. The length of the polarization vectors corresponds to the polarization fraction. The vectors are plotted with $\sim$ 5 segments per beam.}
\label{maps}
\end{figure*}

Figure \ref{maps} shows the Stokes $I$ dust emission at 3~mm from our observations.  The dust emission morphology is consistent with previous observations.   The red line segments are the observed polarization direction, which we will refer to as polarization angle ``vectors'', although they are not true vectors as they have no unique direction.
Table \ref{obs} lists the measured values for total intensity and polarized intensity, as well as the beam size. We also list the 3$\sigma$ upper limit for the two polarization non-detections.  The integrated Stokes $I$ fluxes of DG Tau, Haro 6-13, and MWC 480 are $\sim$30\% lower than those reported by \citet{2011AA...529A.105G} at 2.7 mm using the Plateau de Bure interferometer. The Haro 6-13 flux at 3 mm is consistent with \citet{kwon} at a level of 10\%.
The integrated Stokes $I$ flux of DL Tau is 20\% higher than that reported in \citet{2011AA...529A.105G}, but it is within 5\% of the 2.7 mm flux reported in \citet{kwon}. V892 Tau's Stokes $I$ flux is 20\% lower than that reported in \citet{ricci} at 2.9 mm.  We attribute this variance as likely due to typical absolute amplitude calibration uncertainties with incomplete {\it u,v} coverage in the previous observations.

\begin{table*}[ht]
\centering
{Total and Polarized Intensities Beam Sizes, and Inclination Angles for all Sources} \label{tab:title}
\begin{center}
\begin{tabular}{ |c|c|c|c|c|c|c| } 
 \hline
 Source & Inc. ($i$) & $I$ flux (mJy) & $\sigma_I$ ($\mu$Jy/beam) & $P$ peak ($\mu$Jy/beam) & $\sigma_P$ ($\mu$Jy/beam) & Beam Size\\
 \hline
 DG Tau &32$^{\circ}$ & $42.71 \pm 0.64$ & 16.0 & 127 & 15 & 0$\farcs$23$\times$0$\farcs$22\\ 
 \hline
 Haro 6-13 &40$^\circ$.& $24.96 \pm 0.08$ & 19.2 &129 & 16  & 0$\farcs$27$\times$0$\farcs$22\\ 
 \hline
 MWC 480 &36$^\circ$& $23.21 \pm 0.41$ &17.7 &96.2 & 15.8 & 0$\farcs$33$\times$0$\farcs$26\\
 \hline
 RY Tau &65$^\circ$& $29.50 \pm 0.09$ & 29.7 &243 &19  & 0$\farcs$43$\times$0$\farcs$30\\
 \hline
 V892 Tau &59$^\circ$& $44.22 \pm 0.22$ & 28.1&\textless 128 &43 & 0$\farcs$26$\times$0$\farcs$20\\
 \hline
 DL Tau &45$^\circ$& $32.90 \pm 0.36$ &23.5 & \textless 124 & 42& 0$\farcs$24$\times$0$\farcs$19\\
 \hline
\end{tabular}
\caption{Total and polarized intensities beam sizes, and inclination angles for all sources. 3$\sigma$ upper limits on polarized intensity are given for V892 Tau and DL Tau. Inclination angles for V892 Tau and DL Tau from \citet{hamidouce10} and \citet{2018arXiv181006044L}, respectively.}
\end{center}
\label{obs}
\end{table*}

The polarization emission morphology from the four disks in Figure \ref{maps} can be qualitatively grouped into two categories: those with azimuthal polarization vectors and two non-polarized ``holes" near the center of the disk (DG Tau and Haro 6-13), and those with polarization vectors parallel to the disk minor axis and polarized emission only near the center of the disk (RY Tau and MWC 480).

\begin{figure*}[ht!]
\gridline{\fig{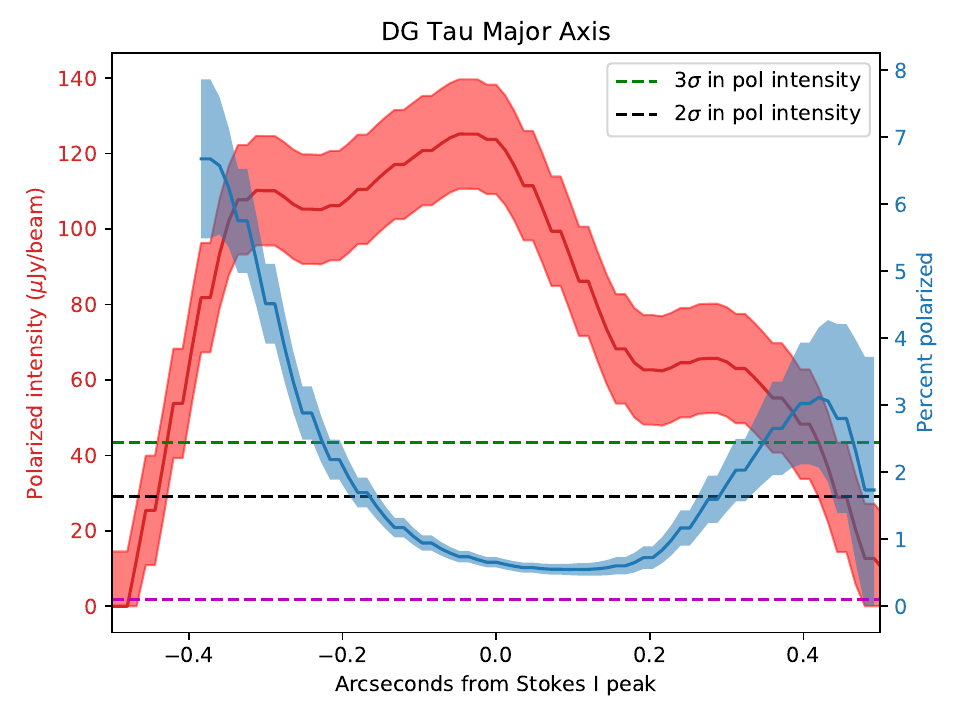}{0.4\textwidth}{(a)}
          \fig{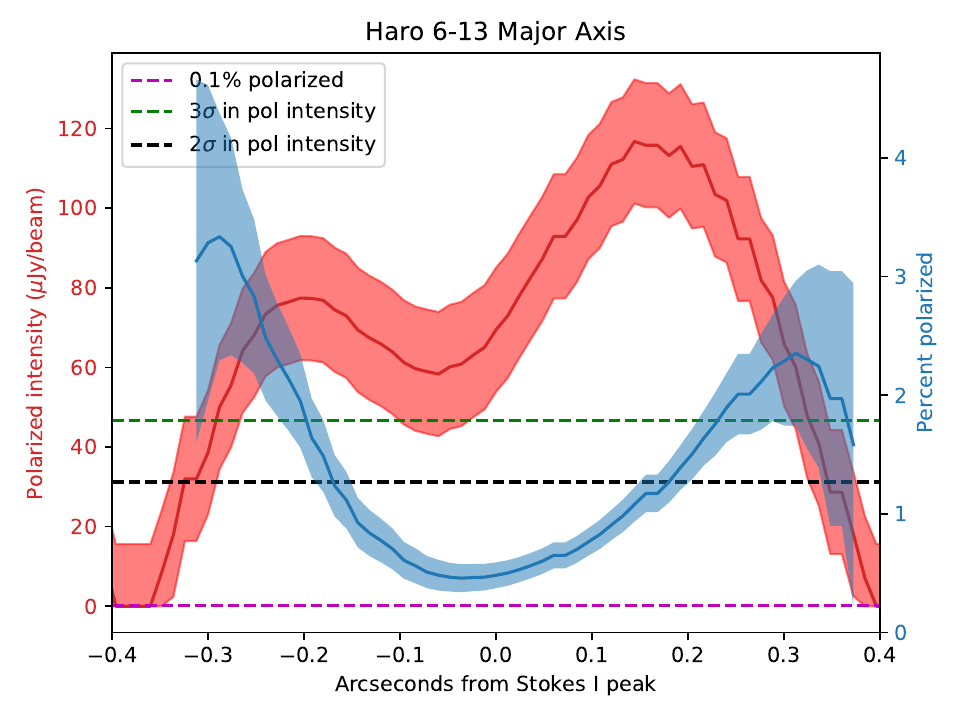}{0.4\textwidth}{(b)}}
\gridline{\fig{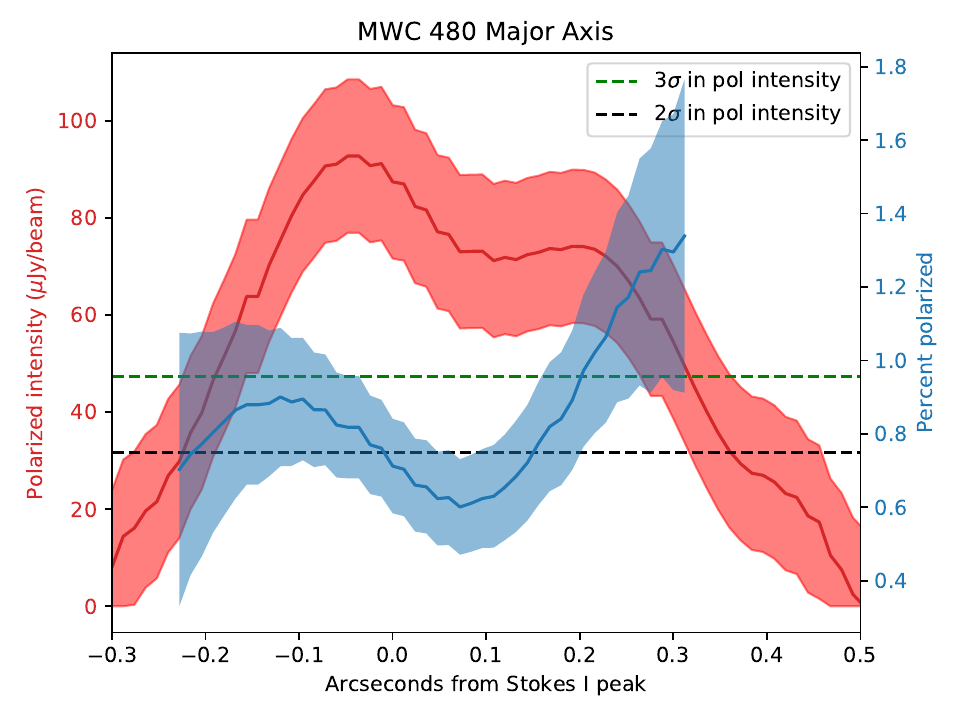}{0.4\textwidth}{(c)}
          \fig{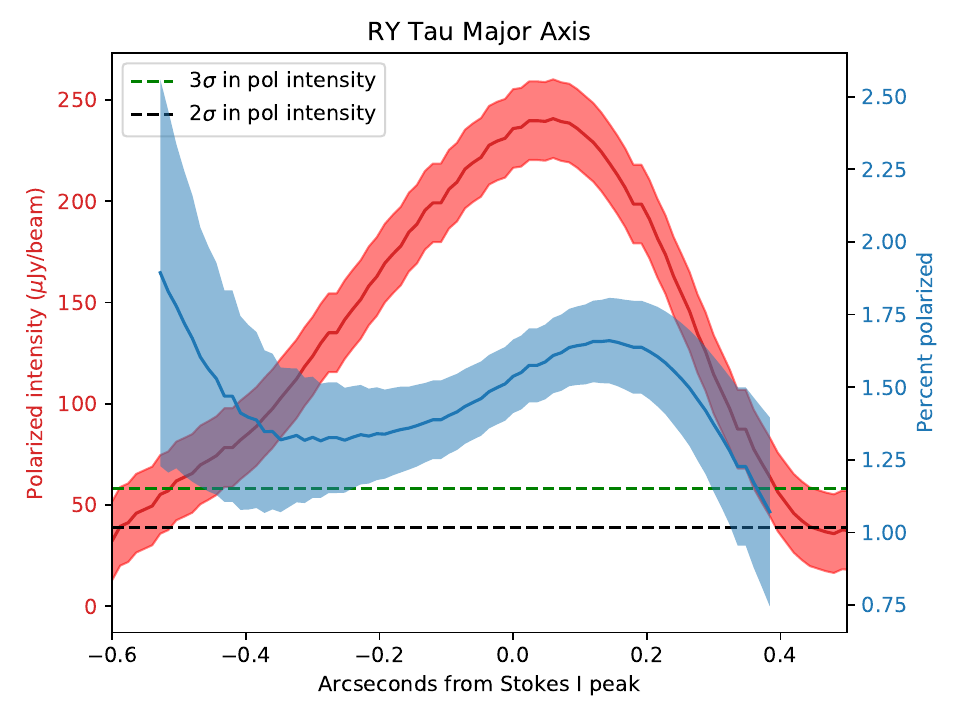}{0.4\textwidth}{(d)}}
\caption{Polarized intensity and percent polarization vs. distance from Stokes I peak along the major axes of the disks. The shaded areas represent $\pm 1\sigma$ statistical uncertainties. Positive distances correspond to north and west of the Stokes $I$ peak for DG Tau, Haro 6-13, and MWC 480, and south and west of the Stokes $I$ peak or RY Tau.}
\label{polint1}
\end{figure*}

\begin{figure*}[ht!]
\gridline{\fig{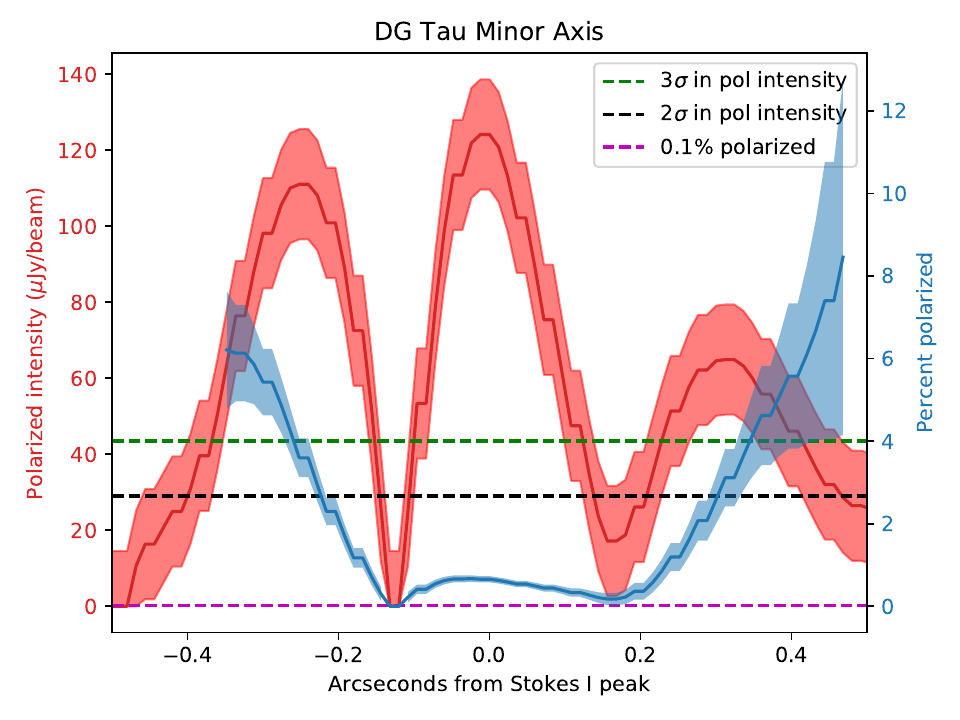}{0.4\textwidth}{(a)}
          \fig{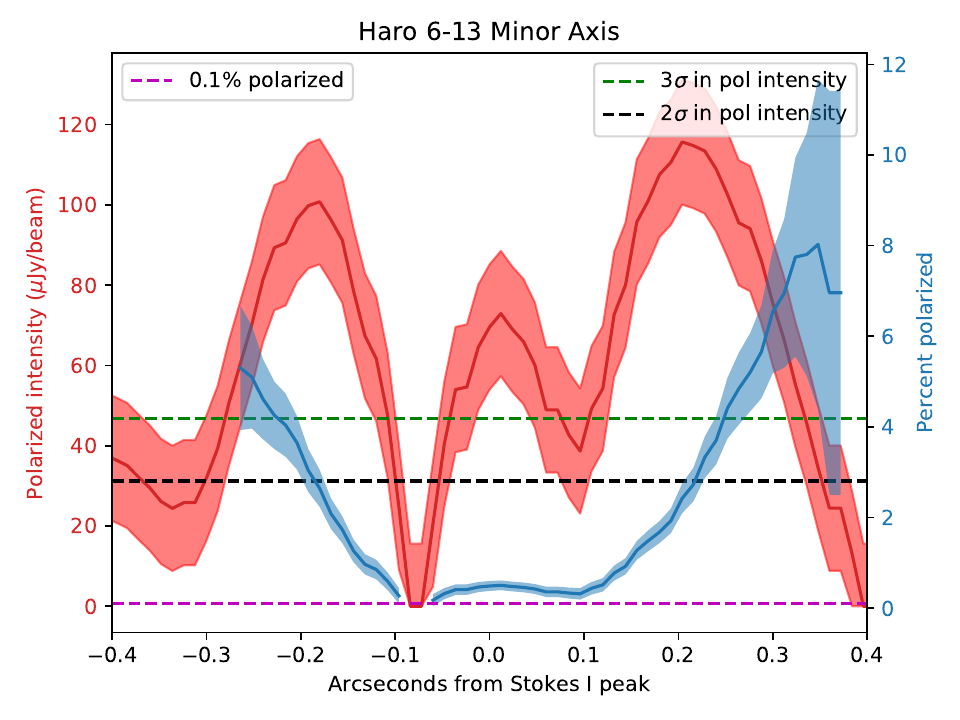}{0.4\textwidth}{(b)}}
\gridline{\fig{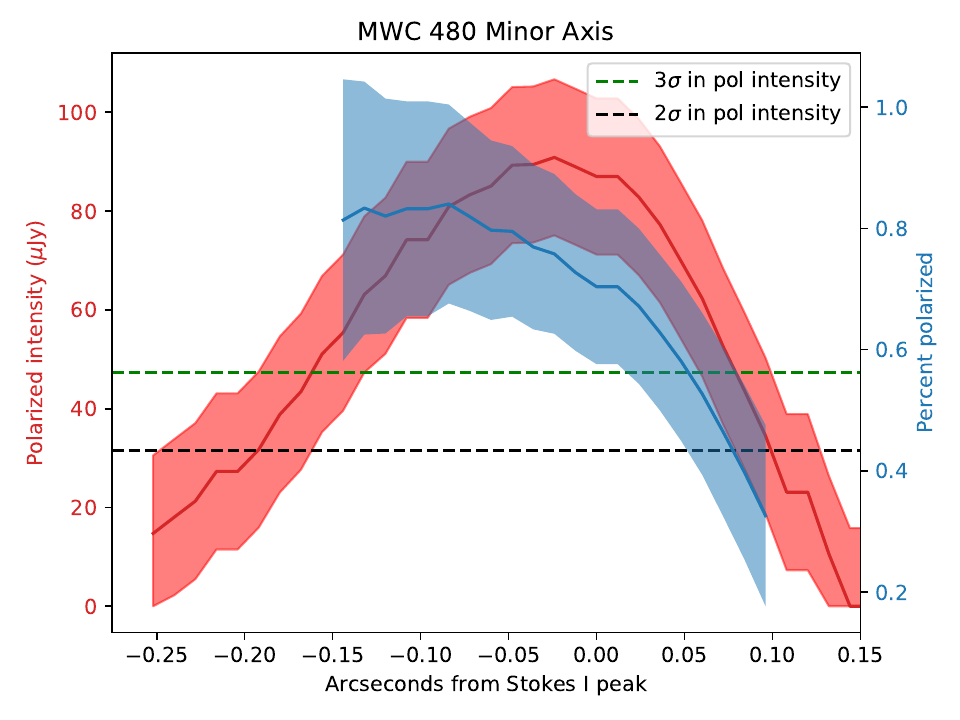}{0.4\textwidth}{(c)}
          \fig{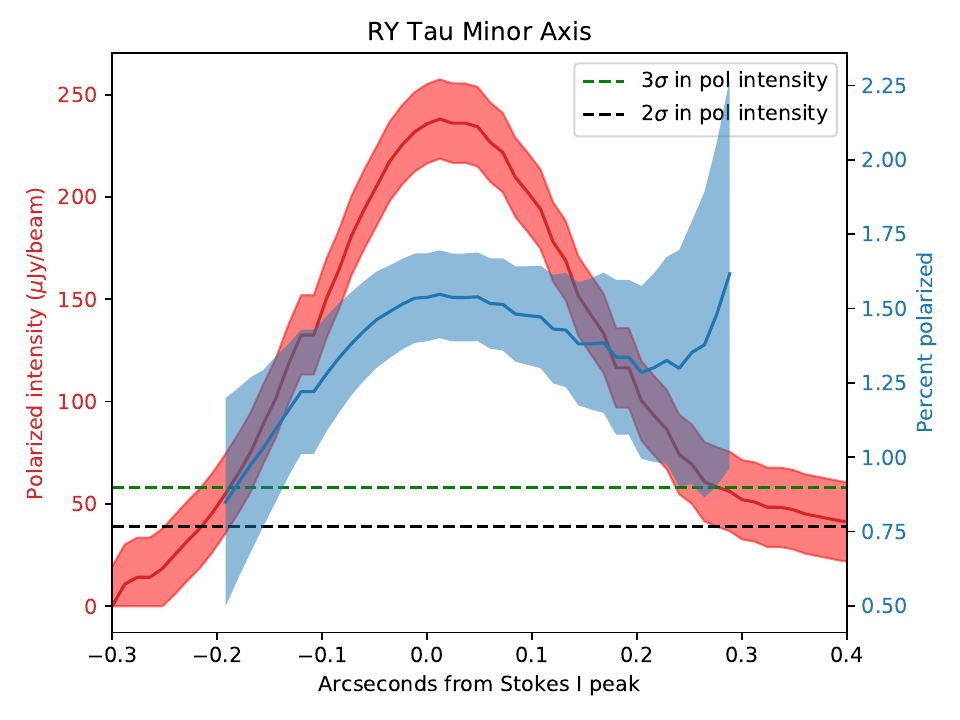}{0.4\textwidth}{(d)}}
\caption{Polarized intensity and percent polarization vs. distance from Stokes I peak along the projected minor axes of the disks. The shaded areas represent $\pm 1 \sigma$ statistical uncertainties. Errors on polarization fraction are only shown in regions where $P > 0$. Positive distances correspond to south and west of the Stokes $I$ peak for DG Tau, Haro 6-13, and MWC 480, and north and west of the Stokes $I$ peak for RY Tau.}
\label{polint2}
\end{figure*}

The two types of polarization morphologies can also be seen by plotting the polarized intensity 
with distance from the Stokes I peak along the major and minor axes of the disks (Figures \ref{polint1} and \ref{polint2}). 
DG Tau and Haro 6-13 in Figure \ref{polint2} both have two low-polarization ``holes'' along the minor axis. These depolarized regions may be places where the polarization orientation changes rapidly within one beamwidth. Haro 6-13 also exhibits a more pronounced asymmetry along its major axis than DG Tau.  On the other hand, the polarized intensities along the major and minor axes of RY Tau and MWC 480 in Figure \ref{polint1} and \ref{polint2} peak near the Stokes I peak of the sources, and these sources lack the low-polarization holes seen in DG Tau and Haro 6-13.

DG Tau and Haro 6-13 have higher fractional polarizations than MWC 480 and RY Tau. In DG Tau and Haro 6-13, the polarization fraction is highest near the edges of the polarized region. The low polarization fraction near the centers of these disks may be a beam dilution effect; a polarization pattern that varies azimuthally within one beam will be averaged down to a lower apparent polarization fraction. We note that if polarized emission were present near the edges of the disk MWC 480 and RY Tau at polarization fractions similar to those in Haro 6-13 and DG Tau, it would have been detected in these observations. The distribution of polarized emission along the major axis of Haro 6-13 is also noticeably asymmetrical (see Figure \ref{polint1}(b)), while the disk is symmetrical in Stokes I. This asymmetry rises to about the 2$\sigma$ level. If observations with higher signal-to-noise ratios confirm that this feature is real, it will warrant further investigation.
The polarized regions of RY Tau and MWC 480 are only about a beam across. This limits the scope of the conclusions we can draw about these sources, since we only have a small number of independent measurements of the polarization angle across the disk.

\section{Discussion} \label{sec:discussion}

The polarization morphology observed in Haro 6-13 and DG Tau at 3 mm resembles that observed in HL Tau at the same wavelength. RY Tau and MWC 480, however, exhibit polarization parallel to the disk minor axis, similar to HL Tau at 870 $\mu$m.  This comparison is still true regarding the way polarized intensity changes across the disks. 
These differences at the same wavelength for the two groups of sources implies different polarization mechanisms for very similar sources. 

To explore if these differences could be attributed to inclination or signal-to-noise, we compared the observed polarization patterns to those predicted from a simple model. The model creates a map of the polarized intensity and polarization angle based on the disk's inclination and position angle, as well as the beam size and position angle, for one of four mechanisms: Rayleigh scattering, radiative alignment, mechanical alignment through the Gold mechanism \citep[as described in][]{gold52}, and alignment to a toroidal magnetic field. In the radiative model, the polarization angle is calculated by rotating the radial direction in the disk plane by 90$^\circ$. The polarization angle in the case of mechanical alignment through the Gold mechanism is the position angle of the toroidal direction in the sky plane, and the polarization angle in the case of magnetic alignment is perpendicular to the toroidal direction. In the case of self-scattering, the model sets the polarization angle parallel to the disk's minor axis. The polarization fraction produced by the three alignment mechanisms (radiative, magnetic, and mechanical) depends on the cosine of the angle between the dust grain alignment axis and the line of sight direction. The polarization fraction produced by self-scattering depends on the disk inclination angle $i$ as $\sin^2(i)$. The polarization fractions are then multiplied by a simple model for the disk brightness to give the polarized intensity. Because the models are inherently axisymmetric, they cannot explain any of the asymmetry observed in polarized intensity.

\begin{figure*}[ht!]
\gridline{\fig{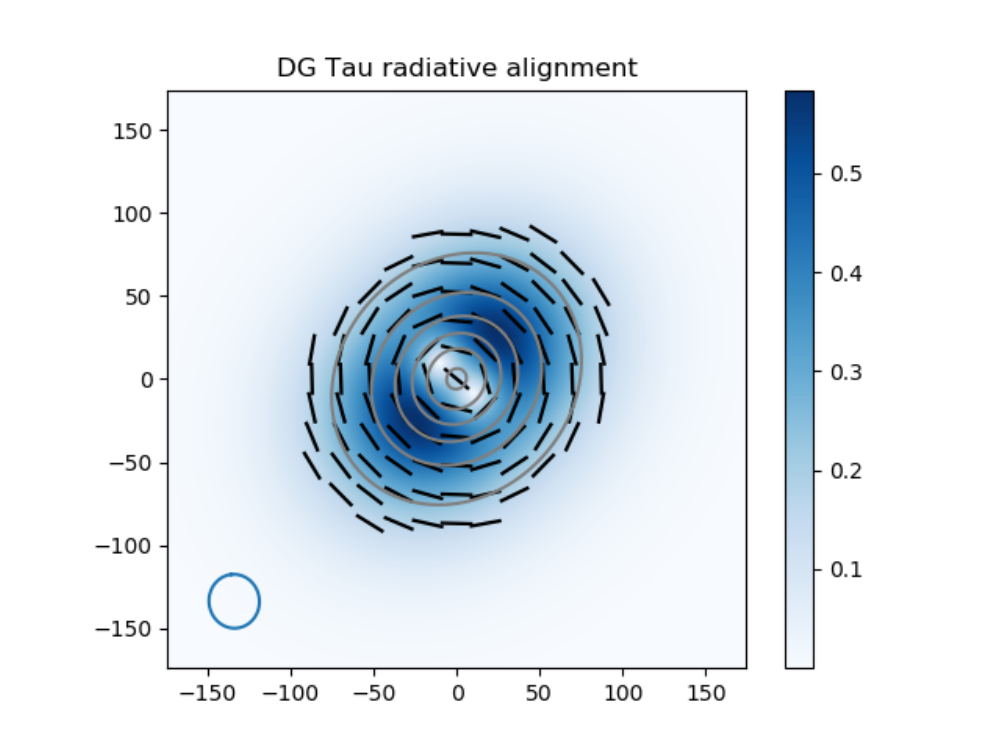}{0.45\textwidth}{(a)}
          \fig{DG_Tauri_composite_newscale.pdf}{0.39\textwidth}{(b)}}
\gridline{\fig{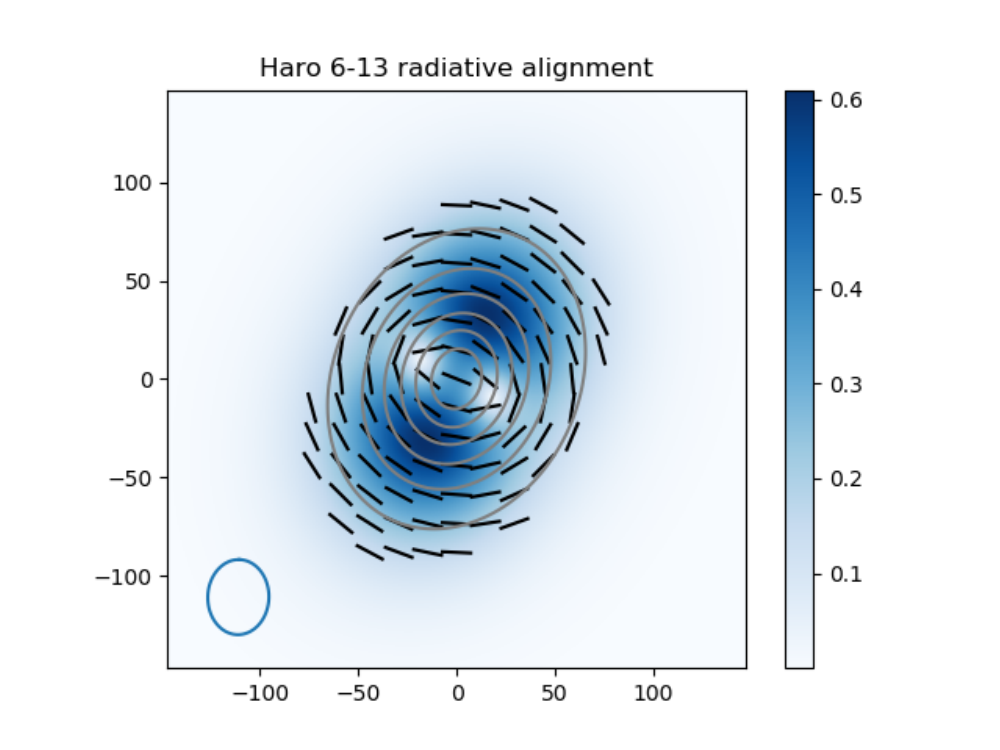}{0.45\textwidth}{(c)}
          \fig{Haro_6-13_composite_scalebar.pdf}{0.39\textwidth}{(d)}}
\caption{Models of expected polarization morphology in DG Tau and Haro 6-13 from radiative alignment \deleted{and MWC 480 and RY Tau from scattering}, alongside the maps from Figure 1. In models, gray contours represent Stokes I, blue shading represents polarized intensity, and black pseudo-vectors represent polarization angle. The color scale is relative and not meant to quantitatively predict polarized intensity values.}
\label{dg_haro_mod}
\end{figure*}

\begin{figure*}
\gridline{\fig{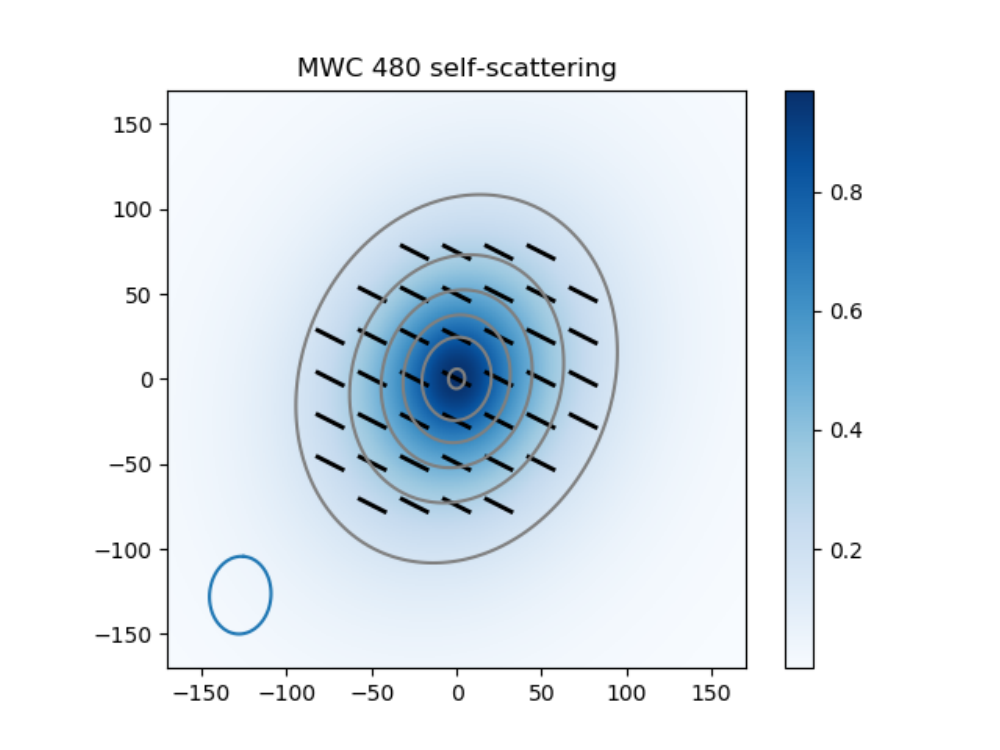}{0.45\textwidth}{(a)}
          \fig{MWC480_composite_scalebar.pdf}{0.39\textwidth}{(b)}}
\gridline{\fig{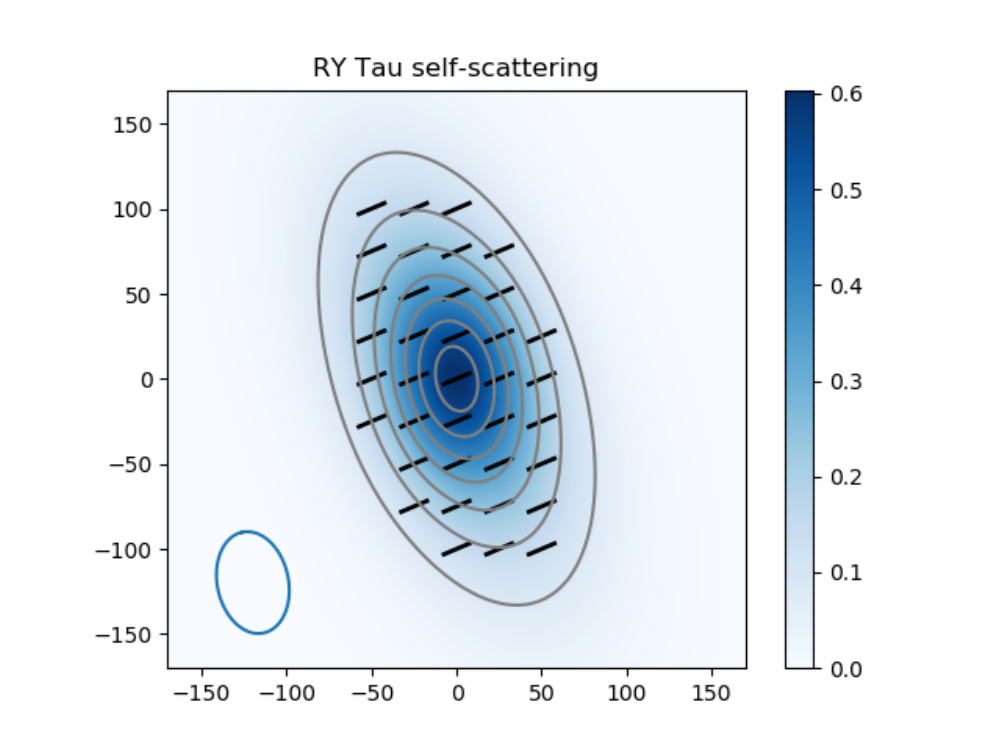}{0.45\textwidth}{(c)}
          \fig{RY_Tauri_scalebar.pdf}{0.39\textwidth}{(d)}}
\caption{Models of expected polarization morphology in MWC 480 and RY Tau from scattering, alongside the maps from Figure 1. In models, gray contours represent Stokes I, blue shading represents polarized intensity, and black pseudo-vectors represent polarization angle. The color scale is relative and not meant to quantitatively predict polarized intensity values.}
\label{mwc_ry_mod}
\end{figure*}

Figures \ref{dg_haro_mod} and \ref{mwc_ry_mod} show the models that best fit the observed polarization for each source, alongside the maps from Figure \ref{maps}. Comparison with our simple model shows that the polarization in DG Tau and Haro 6-13 is broadly consistent with that expected from grain alignment to the radiation anisotropy; the polarization vectors are azimuthally oriented, with two low-polarization holes caused by beam averaging on either side of the disk major axis. In contrast, models of the expected polarization from mechanical alignment through the Gold mechanism for these disks produced azimuthally oriented polarization vectors with low-polarization holes on either side of the disks' minor axes. However, the azimuthal variations in polarized intensity expected from radiative and aerodynamic alignment are not seen in DG Tau and Haro 6-13 (see \citet{yang18}, Figure 12.).  The polarization in RY Tau and MWC 480 is broadly consistent with that expected from scattering; the polarization vectors are aligned with the minor axis of the disk, and the polarized intensity peaks at the center of the disk. In contrast, the model for radiative alignment predicts that polarized emission would peak at two points along the major axis in RY Tau and MWC 480. Additionally, we note that while Haro 6-13 and MWC 480 have nearly the same inclination angle (40$^\circ$ and 36$^\circ$, respectively) they have different polarization morphologies, which indicates that the differences in these two disks cannot be attributed solely to differences in inclination angle.  

The variation in polarization with wavelength in HL Tau has been explained by scattering dominating at 870 $\mu$m and radiation alignment dominating at 3 mm, with a combination of the two mechanisms present at 1.3 mm \citep{2017ApJ...851...55S}. With our observations of DG Tau, Haro 6-13, MWC 480, and RY Tau, we provide the first evidence of different polarization morphologies in otherwise similar disks at 3 mm, implying that different polarization mechanisms may dominate in these disks at the same wavelength.

\subsection{Potential Evolutionary effects}

The differences in polarization mechanisms in these disks may indicate that the disks are at different stages of evolution. Polarization from scattering is present at a longer wavelength in MWC 480 and RY Tau than in HL Tau. This could indicate that MWC 480 and RY Tau have (compared to HL Tau) larger dust grains, which could imply a more evolved disk with time to allow dust to grow to larger sizes. 
To determine whether evolutionary effects are responsible for the variation seen in these disks, we will need observations at other wavelengths to determine more quantitatively where the transitions between polarization patterns take place.

The dust opacity spectral index ($\beta$) has been used to estimate grain sizes in circumstellar disks with assumed dust properties. \citet{kwon} obtained $\beta$ values of 0.6745 $\pm$ 0.0069 (viscous accretion disk model) to 0.615 $\pm$ 0.006 (power-law disk model) for HL Tau and $\beta$ values consistent with zero or a small positive number for Haro 6-13. Even with the uncertainty of 0.25 on these values, HL Tau's $\beta$ value is higher than Haro 6-13's, which is consistent with Haro 6-13 being a more evolved disk. The similar polarization morphologies, on the other hand, imply that the two disks have similar grain sizes. However, there is an order of magnitude discrepancy between the grain size estimates from Rayleigh scattering and those from $\beta$ in HL Tau; scattering gives an estimated maximum grain size of up to $\sim$150 $\mu$m, while $\beta$ gives a maximum grain size of $\sim$1 mm \citep{kataoka16}. The cause of this discrepancy is currently unknown. 

\subsection{Problems with the radiative alignment model}

Although HL Tau's 3~mm polarization morphology has been used as an example of radiation alignment, it is not well fit by this mechanism \citep{yang18}.  The polarization pattern of HL Tau is very similar to DG Tau and HH 6-13 in Figure \ref{maps}, which is an elliptical polarization pattern.  However, radiatively aligned grains produce an intrinsically circular polarization pattern even for inclined disks such as HL Tau. 
To recreate the elliptical pattern, 
\citet{yang18} used 
the aerodynamic alignment mechanism \citep[e.g.,][]{gold52}, which can produce an elliptical  pattern if the dust grains are aligned aerodynamically by the difference in rotation speed between the dust and gas.  Unfortunately, aerodynamic alignment, just like radiation alignment, creates large azimuthal variation in polarized intensity \citep{yang18} that is not seen in HL Tau nor DG Tau and HH 6-13.  In other words, we do not currently have a robust mechanism for polarization in these cases.  Multi-wavelength observations coupled with more complete models that include 3D descriptions of the radiation and disk will be necessary to constrain the mechanism or mechanisms responsible for polarized emission in these disks.

\section{Conclusions}

These observations represent the largest survey of protoplanetary disks in polarization at 3 mm. We find that the polarization morphologies can be qualitatively divided into two categories: those with polarization angles oriented azimuthally in the outer part of the polarized region (DG Tau and Haro 6-13), and those with polarized intensity that peaks at the center of the disk with the angle of polarization parallel to the disk minor axis (RY Tau and MWC 480). We argue that preliminary modeling indicates that these differences do not arise solely from inclination nor from signal-to-noise effects. The differences in polarization morphology may indicate that different polarization mechanisms dominate in different disks at the same observing wavelength. Multi-wavelength observations and more complete modeling, taking into account optical depth effects, disk thickness, and combinations of mechanisms, will be needed to gain a fuller understanding of the processes creating polarized emission in these disks. 

\label{sec:conclusions}

%

\vspace{5mm}
\facilities{ALMA}


\software{CASA (McMullin et al. 2007), 
          astropy (The Astropy Collaboration 2013, 2018), 
          APLpy (Robitaille \& Bressert 2012)
          }


\section*{Acknowledgements}
This paper makes use of the following ALMA data: ALMA \#2017.1.00470.S. ALMA is a partnership of ESO (representing its member states), NSF (USA) and NINS (Japan), together with NRC (Canada), MOST and ASIAA (Taiwan), and KASI (Republic of Korea), in cooperation with the Republic of Chile. The Joint ALMA Observatory is operated by ESO, AUI/NRAO and NAOJ. The National Radio Astronomy Observatory is a facility of the National Science Foundation operated under cooperative agreement by Associated Universities, Inc. REH is supported by an ALMA Student Observing Support Grant. ZYL is supported in part by NASA grant 80NSSCK1095 and NNX14B38G and NSF grant AST-1815785 and 1716259. Woojin Kwon was supported by Basic Science Research Program through the National Research Foundation of Korea (NRF-2016R1C1B2013642). We gratefully appreciate the comments from the editors and anonymous referees that significantly improved this Letter. We would like to thank the NAASC Data Analysts for performing the initial calibration and imaging of this data.

\end{document}